\documentclass{article}
\usepackage{spconf,amsmath,amssymb,graphicx,hyperref}
\hypersetup{hidelinks}
\usepackage{array}
\usepackage{tabularx}
\usepackage{xurl}
\usepackage{microtype}
\usepackage[caption=false,font=small]{subfig}
\usepackage{xcolor}

\title{Not All Relations Are Equal: Relation-Balanced and Calibrated Graph Learning for Provenance-Based Intrusion Detection}

\name{Lijie Zheng$^{1}$, Ji He$^{1}$, Zhiwei Zhang$^{1}$, Alessandro Brighente$^{2}$, Yulong Shen$^{1}$, Mauro Conti$^{2,3}$}
\address{$^{1}$School of Computer Science and Technology, Xidian University, Xi'an, China\\
  $^{2}$Department of Mathematics, University of Padova, Padova, Italy\\
  $^{3}$School of Science and Technology, \"Orebro University, \"Orebro, Sweden}

\begin{document}
\ninept
\maketitle

\begin{abstract}
Provenance-Based Intrusion Detection Systems (PIDSs) detect Advanced Persistent Threats (APTs) by analyzing system interactions. However, existing methods largely treat relations uniformly, overlooking statistical heterogeneity; in CADETS, relation frequencies differ by approximately $140{,}000\times$. This may cause PIDSs to focus more on frequent relations and overlook differences in normal error levels across relations, increasing the risk of false alarms and missed detections. We present RECAL, an unsupervised framework using relation-balanced masked graph learning to better capture rare interaction patterns. It further calibrates reconstruction errors against each relation's benign error distribution to produce comparable anomaly evidence, helping distinguish attacks from benign behavior and reduce false alarms. On three DARPA E3 datasets, RECAL achieves F1 scores of 99.99\%, 99.93\%, and 99.99\%, outperforming the best baseline on each dataset by 0.88, 0.82, and 0.42 percentage points, respectively. Compared with the baseline reporting the lowest FPR, RECAL reduces mean FPR by approximately $105\times$, $4\times$, and $41\times$.
\end{abstract}


\section{Introduction}
\label{sec:intro}

An Advanced Persistent Threat (APT) blends stealthy malicious actions into normal system activity over extended periods, posing challenges for signature-based or rule-based detection~\cite{zipperle2023}. Kernel audit logs record interactions between system entities, and provenance graphs constructed from these logs are widely used to detect such attacks~\cite{hossain2017,milajerdi2019}. Using these provenance graphs, some Provenance-Based Intrusion Detection Systems (PIDSs) apply graph learning to model benign behavior and identify deviations to detect potential attacks~\cite{threatrace,flash,magic,kairos,yang2026provfusion}.

\begin{figure}[htb]
\centering
\includegraphics[width=\linewidth]{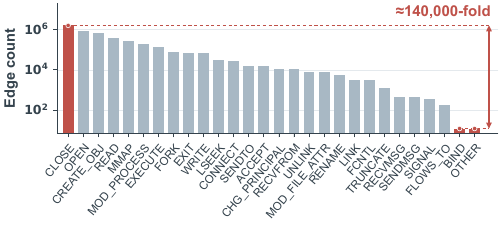}
\caption{CADETS benign training graphs. The x-axis shows audit-log relations; the y-axis shows event counts (log scale).}
\label{fig:obs}
\end{figure}

However, these PIDSs are largely insensitive to the heterogeneity of interaction relations. Across relation types (system-call semantics such as \texttt{read}, \texttt{write}, and \texttt{execute}), audit data exhibits drastically different statistics. As shown in Fig.~\ref{fig:obs}, event counts across interaction relations in the CADETS benign training graphs differ by approximately $140{,}000\times$. Existing methods nevertheless treat all relations uniformly in both learning and judgment, even when some encode relation types as input features~\cite{kairos,magic}. This insensitivity raises two problems. \textbf{(1) Capacity misallocation.} The learning signal is dominated by high-frequency relations, so rare but security-critical interactions, such as payload execution and privilege escalation, may be insufficiently modeled~\cite{threatrace,flash,magic}. \textbf{(2) Cross-relation score miscalibration.} When benign score distributions differ across relations, raw scores do not indicate comparable degrees of deviation~\cite{kriegel2011}. A single global threshold may then over-alert on high-score relations and overlook anomalous behavior on low-score relations~\cite{kairos,unicorn,nodoze}.

To address these problems, we present RECAL, an unsupervised, relation-calibrated PIDS framework that models both frequent and rare interactions through relation-balanced masked graph learning and converts within-relation behavioral deviations into comparable anomaly evidence across relations, thereby reducing false positive rates. The main contributions of this work are summarized as follows.
\begin{itemize}
\item We identify capacity misallocation and cross-relation score miscalibration as two challenges arising from relation heterogeneity in existing PIDSs.
\item We present RECAL, integrating relation-balanced masked graph learning with relation-calibrated detection. Relation-stratified masking, independent decoder heads, and balanced reconstruction mitigate frequent-relation dominance. Empirical quantile calibration and Fisher fusion align and aggregate cross-relation evidence, improving detection and reducing false alarms.
\item We systematically evaluate RECAL on three DARPA E3 datasets. The results confirm its effectiveness, with the highest F1 and lower false positive rates among compared methods. Our implementation is available at \url{https://github.com/Jiex2001/RECAL} to support further research.
\end{itemize}

\section{Methodology}
\label{sec:method}

\begin{figure*}[t]
\centering
\includegraphics[width=0.955\textwidth]{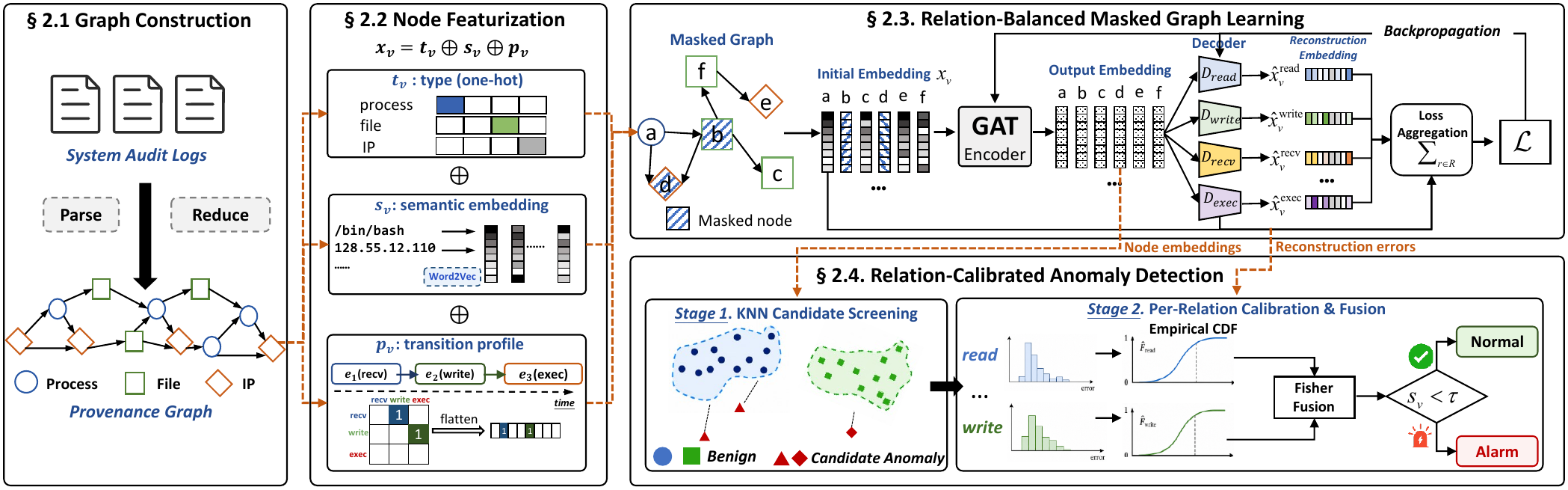}
\caption{Overview of RECAL. Graph construction and node featurization transform audit logs into attributed provenance graphs for relation-balanced masked graph learning. Node embeddings support KNN candidate screening, while relation-specific reconstruction errors support candidate calibration and evidence fusion for final detection. Orange dashed arrows indicate data flow between modules.}
\label{fig:framework}
\end{figure*}

The RECAL pipeline comprises four stages, as shown in Fig.~\ref{fig:framework}. First, graph construction (\S\ref{sec:graph}) converts kernel audit logs into a typed provenance graph. Second, node featurization (\S\ref{sec:feat}) encodes entity types, attribute semantics, and temporal relation transitions as node features. Third, relation-balanced masked graph learning (\S\ref{sec:m1}) improves rare-relation modeling and produces per-relation reconstruction errors. Finally, relation-calibrated anomaly detection (\S\ref{sec:m2}) combines K-Nearest Neighbor (KNN) candidate screening with relation calibration and evidence fusion for two-stage node-level detection.

\subsection{Graph Construction}
\label{sec:graph}

We convert kernel audit logs into a directed heterogeneous provenance graph $G=(V,E,R)$, where $V$ denotes the set of system entities, including processes, files, and network flows; $E$ denotes the set of interaction edges; and $R$ denotes the relation vocabulary defined by the audit logs, including \texttt{read}, \texttt{write}, \texttt{execute}, and \texttt{connect}. During construction, we apply standard provenance graph reduction techniques~\cite{xu2016cpr} to control graph size while preserving causal structure.

\subsection{Node Featurization}
\label{sec:feat}

For each node $v$, the initial feature is $x_v = t_v \oplus s_v \oplus p_v$,
where $T$ is the set of entity types, $t_v\in\{0,1\}^{|T|}$ is the one-hot encoding of the entity type, and $s_v\in\mathbb{R}^{d}$ is a semantic embedding, obtained by tokenizing the attribute strings of the node (process names, file paths, IP addresses) and averaging their Word2Vec vectors~\cite{mikolov2013}. The third part, $p_v\in\mathbb{R}^{|R|^2}$, is a temporal relation-transition profile. Let $B_v=(e_1,\dots,e_L)$ be the $L$ most recent events involving $v$ in ascending temporal order, and let $r(e_i)$ and $\theta(e_i)$ denote the relation type and the timestamp of event $e_i$. The profile entry for a transition $(r,r')$ is
\begin{equation}
p_v[r,r'] = \sum_{i=2}^{L} \mathbb{I}\big[r(e_{i-1})=r \wedge r(e_i)=r'\big]\, e^{-\lambda\,(\theta_v-\theta(e_i))},
\label{eq:profile}
\end{equation}
where $\mathbb{I}[\cdot]$ is the indicator function, $\theta_v$ is the timestamp of the most recent event involving $v$, and $\lambda$ controls the temporal decay. The profile gives each interaction a short-range behavioral context. When the transitions \texttt{receive}$\rightarrow$\texttt{write} and \texttt{write}$\rightarrow$\texttt{execute} are active together, a download-then-execute pattern becomes visible at the feature level. Maintaining the profile costs an $O(L)$ sliding buffer per node and requires no training.

\subsection{Relation-Balanced Masked Graph Learning}
\label{sec:m1}

This module learns normal interaction patterns through masked graph autoencoding~\cite{graphmae}, using node features from benign training graphs as reconstruction targets without attack labels. To mitigate the dominance of frequent relations, we organize reconstruction into relation-specific learning tasks and balance their contributions through masking and reconstruction optimization.

Masking selects the nodes whose features will be reconstructed. To increase reconstruction opportunities for rare relations, we assign each relation observed in the training data a masking rate based on its event frequency,
\begin{equation}
p_r = \mathrm{clip}\big(p_0\,(\bar{f}/f_r)^{\gamma},\; p_{\min},\; p_{\max}\big),
\label{eq:maskrate}
\end{equation}
where $f_r$ is the training event count of relation $r$, $\bar{f}$ is the median nonzero frequency, $p_0$ is the base masking rate, and $p_{\min}$ and $p_{\max}$ are its bounds. The exponent $\gamma$ controls the emphasis on rare relations, with $\gamma=0$ giving equal per-relation rates. For each relation $r$, we independently sample its participating nodes with probability $p_r$; a node is masked if selected under any relation. Its overall masking probability therefore depends on both relation frequencies and the number of participating relations. We replace the input features of selected nodes with a learnable mask vector.

On the resulting masked graph, the encoder uses edge-type-conditioned attention~\cite{velickovic2018,rgatconv} to produce node embeddings. We then average the embeddings of node $v$'s neighbors connected by relation $r$, including both incoming and outgoing neighbors, to obtain $m_v^r$. Each relation has an independent lightweight linear decoder $D_r$ that reconstructs the masked node features as $\hat{x}_v^r = D_r(m_v^r)$. This allows different interaction patterns to be modeled separately using shared node representations.

Because the number of reconstruction samples still varies across relations, we average the loss within each relation and sum these averages with equal weight,
\begin{equation}
\mathcal{L} = \sum_{\substack{r\in R,|M_r|>0}}\frac{1}{|M_r|}\sum_{v\in M_r}\ell\big(\hat{x}_v^r,\, x_v\big),
\label{eq:loss}
\end{equation}
where $M_r$ is the set of masked nodes participating in relation $r$ and $\ell$ is the scaled cosine error~\cite{graphmae}. The objective weights each nonempty relation's mean reconstruction loss equally, reducing the bias due to unequal sample counts. Per-relation reconstruction also retains errors associated with different interactions, allowing subsequent detection to interpret these errors against relation-specific benign baselines.

\subsection{Relation-Calibrated Anomaly Detection}
\label{sec:m2}

To address cross-relation score miscalibration, we interpret reconstruction errors against each relation's benign error distribution. We reserve the tail of the benign period for calibration and exclude it from masked graph model training. After training, we obtain reconstruction errors on the calibration graph through batched masking and estimate a relation-specific empirical Cumulative Distribution Function (CDF) $\hat{F}_r$~\cite{li2023ecod}.

During the detection stage, a KNN detector~\cite{magic,angiulli2002} measures the average nearest-neighbor distance between node embeddings and a benign reference set. A deliberately relaxed threshold yields a high-recall candidate set. For each candidate node $v$, we map its reconstruction error $e_v^r$ to a within-relation benign quantile $q_v^r = \hat{F}_r(e_v^r)$ using the corresponding reference distribution.
This transformation places errors from different relations on a common quantile scale, allowing their deviations from normal behavior to be compared.

We aggregate anomaly evidence across the relations in which each candidate node participates. Let $R(v)$ denote the set of relation types involving $v$ and $k_v=|R(v)|$ their number. We interpret $p_v^r = 1-q_v^r$ as an empirical upper-tail probability estimate and use Fisher's method~\cite{fisher1932} to fuse the evidence,
\begin{equation}
s_v = F_{\chi^2_{2k_v}}\Big(-2\textstyle\sum_{r\in R(v)}\ln p_v^r\Big),
\label{eq:score}
\end{equation}
where $F_{\chi^2_{2k_v}}$ is the CDF of the $\chi^2$ distribution with $2k_v$ degrees of freedom. The score accounts for the number of participating relations and accumulates anomaly evidence across them. Since relation-specific errors share node embeddings and may be dependent, the $\chi^2$ distribution serves as an approximate reference. An alarm is raised when $s_v$ exceeds the decision threshold $\tau$.

Reconstruction errors for test nodes are obtained through batched masked inference. Subsequent calibration and fusion require only quantile-table lookups and $\chi^2$ CDF evaluations, adding little computational overhead.

\section{Experiments}
\label{sec:exp}

This section evaluates RECAL's detection performance, component contributions, parameter sensitivity, and computational efficiency.

\subsection{Experimental Setup}
\label{sec:setup}

\begin{table}[t]
\centering
\caption{Statistics of the DARPA E3 scenarios.}
\label{tab:datasets}

\small
\setlength{\tabcolsep}{2.4pt}

\begin{tabular}{l|ccccc}
\hline\hline
\textbf{Dataset} & \textbf{Nodes} & \textbf{Edges} & \textbf{Malicious} & \textbf{Mal.} & \textbf{Size} \\
\hline
CADETS & 1,452,123 & 7,164,818 & 12,857 & 0.89\% & 19.3\,GB \\
THEIA  & 1,558,101 & 3,574,557 & 25,338 & 1.63\% & 18.8\,GB \\
TRACE  & 3,197,278 & 4,661,252 & 68,135 & 2.13\% & 16.2\,GB \\
\hline
Total  & 6,207,502 & 15,400,627 & 106,330 & 1.71\% & 54.3\,GB \\
\hline\hline
\end{tabular}
\end{table}

\noindent\textbf{Datasets.} We evaluate on three scenarios of DARPA TC Engagement 3 (CADETS, THEIA, and TRACE)~\cite{keromytis2018}, whose scale and malicious-node ratios are summarized in Table~\ref{tab:datasets}. Purely benign time spans are used for training, the tail of the benign span is held out as the calibration set (Sec.~\ref{sec:m2}), and time spans containing attacks are used for testing. We adopt the node-level labels of THREATRACE~\cite{threatrace} and follow its evaluation protocol, reporting Precision (Prec.), Recall (Rec.), F1 score (F1), and False Positive Rate (FPR). Inspired by MAGIC's evaluation procedure~\cite{magic}, we perform a linear search over candidate thresholds for each dataset and select $\tau$ at the operating point with the highest test F1.

\noindent\textbf{Baselines.} We select representative baselines and State-of-the-Art (SOTA) methods, including Log2vec~\cite{log2vec}, THREATRACE~\cite{threatrace}, Unicorn~\cite{unicorn}, FLASH~\cite{flash}, MAGIC~\cite{magic}, STGAN~\cite{stgan}, and AEGIS~\cite{aegis}, to evaluate RECAL's detection performance and false positive control. Experiments run on a single NVIDIA RTX 5090 GPU. Our results in tables~\ref{tab:main} and~\ref{tab:ablation} are means over five seeds (0--4); other experiments use seed 0.

\subsection{Main Results}
\label{sec:results}

\begin{table*}[t]
\centering
\caption{Performance Comparison.}
\label{tab:main}

\footnotesize

\setlength{\tabcolsep}{3.4pt}

\begin{tabular}{l|cccc|cccc|cccc}
\noalign{\hrule height 0.8pt}
& \multicolumn{4}{c|}{\textbf{CADETS}} & \multicolumn{4}{c|}{\textbf{THEIA}} & \multicolumn{4}{c}{\textbf{TRACE}} \\
\cline{2-13}
\textbf{System} & \textbf{Prec.}\makebox[0pt][l]{\,$\boldsymbol{\uparrow}$} & \textbf{Rec.}\makebox[0pt][l]{\,$\boldsymbol{\uparrow}$} & \textbf{F1}\makebox[0pt][l]{\,$\boldsymbol{\uparrow}$} & \textbf{FPR}\makebox[0pt][l]{\,$\boldsymbol{\downarrow}$} & \textbf{Prec.}\makebox[0pt][l]{\,$\boldsymbol{\uparrow}$} & \textbf{Rec.}\makebox[0pt][l]{\,$\boldsymbol{\uparrow}$} & \textbf{F1}\makebox[0pt][l]{\,$\boldsymbol{\uparrow}$} & \textbf{FPR}\makebox[0pt][l]{\,$\boldsymbol{\downarrow}$} & \textbf{Prec.}\makebox[0pt][l]{\,$\boldsymbol{\uparrow}$} & \textbf{Rec.}\makebox[0pt][l]{\,$\boldsymbol{\uparrow}$} & \textbf{F1}\makebox[0pt][l]{\,$\boldsymbol{\uparrow}$} & \textbf{FPR}\makebox[0pt][l]{\,$\boldsymbol{\downarrow}$} \\
\hline
Log2vec~\cite{log2vec}       & 49.20\% & 84.59\% & 62.21\% & 1.59\% & 62.49\% & 66.05\% & 64.22\% & 0.29\% & 54.39\% & 78.27\% & 64.18\% & 1.83\% \\
THREATRACE~\cite{threatrace} & 90.42\% & 99.97\% & 94.96\% & 0.19\% & 87.04\% & 99.74\% & 92.96\% & 0.11\% & 71.56\% & 99.99\% & 83.42\% & 1.11\% \\
Unicorn~\cite{unicorn}       & 31.00\% & 100.00\% & 47.00\% & -- & 67.00\% & 67.00\% & 67.00\% & -- & 28.00\% & \textbf{100.00\%} & 34.00\% & -- \\
FLASH~\cite{flash}           & 94.69\% & 99.99\% & 97.27\% & 0.10\% & 93.10\% & 99.83\% & 96.35\% & 0.05\% & 94.65\% & 99.99\% & 97.25\% & 0.16\% \\
MAGIC~\cite{magic}           & 94.40\% & 99.77\% & 97.01\% & 0.22\% & 98.23\% & 99.99\% & 99.11\% & 0.14\% & 99.17\% & 99.98\% & 99.57\% & 0.09\% \\
STGAN~\cite{stgan}           & 98.40\% & 99.83\% & 99.11\% & 0.0426\% & 94.49\% & 99.83\% & 97.09\% & 0.0156\% & 99.50\% & 98.83\% & 99.16\% & 0.0777\% \\
AEGIS~\cite{aegis}           & 100\% & 99\% & 99\% & -- & 96\% & 97\% & 97\% & -- & 99\% & 99\% & 99\% & -- \\
\hline
\textbf{RECAL (ours)} & 99.99\% & 99.99\% & \textbf{99.99\%} & \textbf{0.0004\%} & \textbf{99.95\%} & 99.92\% & \textbf{99.93\%} & \textbf{0.0041\%} & \textbf{99.98\%} & \textbf{100.00\%} & \textbf{99.99\%} & \textbf{0.0019\%} \\
\noalign{\hrule height 0.8pt}
\end{tabular}
\par\vspace{2pt}
\parbox{\textwidth}{\footnotesize\raggedright Arrows indicate whether higher ($\uparrow$) or lower ($\downarrow$) values are preferred; ``--'' denotes unreported values.}
\end{table*}

Table~\ref{tab:main} shows that RECAL achieves the highest F1 among the compared methods on CADETS, THEIA, and TRACE, reaching 99.99\%, 99.93\%, and 99.99\%, respectively. Since recall is nearly saturated for several strong baselines, performance differences mainly lie in false positive control and the resulting precision gains. RECAL models different interaction patterns through per-relation reconstruction, then calibrates and aggregates anomaly evidence against each relation's benign error distribution. This helps reduce the influence of differing error scales across relations on detection decisions and limit false alarms caused by normal behavior. RECAL achieves mean FPRs of 0.0004\%, 0.0041\%, and 0.0019\% on CADETS, THEIA, and TRACE, respectively. Compared with STGAN, the baseline with the lowest reported FPRs in the table, RECAL achieves approximately $105\times$, $4\times$, and $41\times$ reductions in mean FPR on CADETS, THEIA, and TRACE, respectively.

Across all five seeds, 18 nodes on THEIA remain undetected, including 17 associated with \texttt{/home/admin/profile}. Further inspection reveals substantial overlap between their local interaction patterns and benign behavior. Although these nodes enter the candidate set, their calibrated and fused anomaly evidence remains below the final detection threshold. This highlights a limitation of deviation-based detection when the local behavior of attack-associated entities closely resembles normal activity.

\subsection{Ablation Study}
\label{sec:ablation}

\begin{table}[t]
\centering
\caption{Results of Ablation Study (\%).}
\label{tab:ablation}

\footnotesize

\setlength{\tabcolsep}{2.5pt}

\begin{tabularx}{\linewidth}{l|*{2}{>{\centering\arraybackslash\leavevmode}X}|*{2}{>{\centering\arraybackslash\leavevmode}X}|*{2}{>{\centering\arraybackslash\leavevmode}X}}
\noalign{\hrule height 0.8pt}
& \multicolumn{2}{c|}{\textbf{CADETS}} & \multicolumn{2}{c|}{\textbf{THEIA}} & \multicolumn{2}{c}{\textbf{TRACE}} \\
\cline{2-7}
\textbf{Variant} & \textbf{F1}\makebox[0pt][l]{\,$\boldsymbol{\uparrow}$} & \textbf{FPR}\makebox[0pt][l]{\,$\boldsymbol{\downarrow}$} & \textbf{F1}\makebox[0pt][l]{\,$\boldsymbol{\uparrow}$} & \textbf{FPR}\makebox[0pt][l]{\,$\boldsymbol{\downarrow}$} & \textbf{F1}\makebox[0pt][l]{\,$\boldsymbol{\uparrow}$} & \textbf{FPR}\makebox[0pt][l]{\,$\boldsymbol{\downarrow}$} \\
\hline
w/o BL & 99.26 & 0.0569 & 99.94 & 0.0043 & 99.44 & 0.0513 \\
w/o CD & 96.05 & 0.3581 & 99.44 & 0.0838 & 99.51 & 0.0128 \\
z-score calib. & 97.09 & 0.2261 & 99.50 & 0.0741 & 99.86 & 0.0311 \\
max fusion & 99.08 & 0.0699 & 99.48 & 0.0801 & 99.98 & 0.0039 \\
mean fusion & 98.67 & 0.0958 & 95.49 & 0.7652 & 99.57 & 0.0714 \\
\hline
\textbf{RECAL} & \textbf{99.99} & \textbf{0.0004} & 99.93 & \textbf{0.0041} & \textbf{99.99} & \textbf{0.0019} \\
\noalign{\hrule height 0.8pt}
\end{tabularx}
\end{table}

\begin{figure}[t]
\centering
\includegraphics[width=0.3483\linewidth]{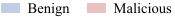}\par
\captionsetup[subfloat]{farskip=0pt,captionskip=2pt}
\subfloat[Before calibration\label{fig:calib-before}]{\includegraphics[width=0.5390\linewidth]{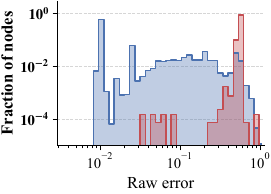}}\hfill
\subfloat[After calibration\label{fig:calib-after}]{\includegraphics[width=0.4569\linewidth]{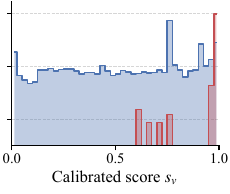}}
\caption{Score distributions of benign and malicious nodes before and after relation-conditional calibration on CADETS.}
\label{fig:calib}
\end{figure}

Table~\ref{tab:ablation} compares five-seed mean performance across learning and detection configurations under the same training budget. The variant using uniform masking and a shared decoder no longer produces per-relation reconstruction errors, so detection falls back to KNN. This joint variant yields lower mean F1 on CADETS and TRACE and slightly higher mean F1 on THEIA, with higher mean FPRs than the full model across all three scenarios. With Relation-Balanced Learning (BL) retained, removing Relation-Calibrated Detection (CD) reduces mean F1 by 3.94, 0.50, and 0.48 percentage points on CADETS, THEIA, and TRACE, respectively, supporting this module's role in both anomaly detection and false positive control. Fig.~\ref{fig:calib} provides a qualitative illustration on CADETS, where benign and malicious node scores are more clearly separated after relation-conditional calibration. Replacing empirical quantile calibration with z-scores also reduces F1 across all three scenarios and raises the mean FPR on CADETS to 0.2261\%, indicating that standardization based only on the mean and variance does not provide the same false positive control. Maximum and mean aggregation both lower F1 and increase FPR. The former retains only the strongest single-relation evidence, while the latter may dilute localized anomalies; Fisher's method accumulates evidence across relations.

\subsection{Parameter Sensitivity}
\label{sec:sensitivity}

\begin{figure}[t]
\centering
\includegraphics[width=\linewidth]{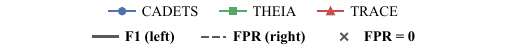}\par

\captionsetup[subfloat]{farskip=0pt,captionskip=2pt}
\subfloat[Calibration-set ratio\label{fig:sens-calib}]{\includegraphics[width=0.495\linewidth]{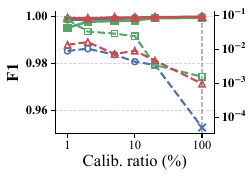}}\hfill
\subfloat[Inference masking rounds $R$\label{fig:sens-rounds}]{\includegraphics[width=0.495\linewidth]{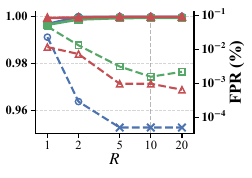}}\par
\subfloat[Masking exponent $\gamma$\label{fig:sens-gamma}]{\includegraphics[width=0.495\linewidth]{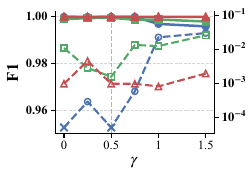}}\hfill
\subfloat[Learning rate\label{fig:sens-lr}]{\includegraphics[width=0.495\linewidth]{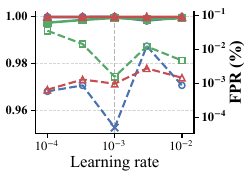}}
\caption{Parameter sensitivity of RECAL. Solid lines show F1 (left axis) and dashed lines show FPR (right axis, log scale).}
\label{fig:sens}
\end{figure}

Fig.~\ref{fig:sens} shows F1 above 99.5\% across all three scenarios with only 1\% of the reserved calibration nodes. Using more benign calibration samples generally helps lower the FPR. Increasing masking rounds preserves more visible neighborhood context, and performance improves as $R$ rises from 1 to 5. Between 5 and 20 rounds, F1 varies by less than 0.01 percentage points per scenario, while forward-pass cost increases. For the masking exponent, $\gamma=0.5$ yields the highest F1 and lowest FPR on CADETS and THEIA; larger values increase their FPR. F1 on TRACE varies by less than 0.02 percentage points. Across tested learning rates from $10^{-4}$ to $10^{-2}$, F1 exceeds 99.7\% in all scenarios, although FPR varies. The default $10^{-3}$ achieves the highest F1 and lowest FPR on CADETS and THEIA, while F1 on TRACE changes little.

\subsection{Efficiency}
\label{sec:efficiency}

\begin{figure}[t]
\centering

\captionsetup[subfloat]{farskip=0pt,captionskip=2pt}
\subfloat[Training and inference time\label{fig:eff-time}]{\includegraphics[width=0.495\linewidth]{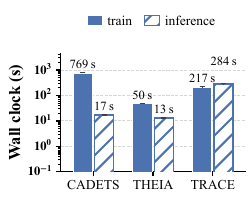}}\hfill
\subfloat[Inference time by stage\label{fig:eff-stages}]{\includegraphics[width=0.495\linewidth]{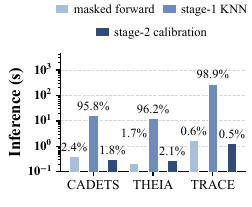}}
\caption{Training and inference costs of RECAL (three-run average). Percentages are each stage's share of total inference time.}
\label{fig:efficiency}
\end{figure}

Fig.~\ref{fig:efficiency} reports the runtime of RECAL. On TRACE, the largest scenario by node count, training for 200 epochs takes about 217 seconds. Inference over approximately 3.3 million nodes across its five test graphs takes about 284 seconds. KNN reference index construction and queries account for 98.9\% of this inference time and are the main computational cost. In contrast, the score calibration and Fisher fusion stage accounts for only approximately 0.5\% to 2.1\% of inference time across all three scenarios, indicating its low computational overhead. The reported inference times exclude preprocessing, data loading, and one-time calibration reference construction.

\section{Conclusion}
\label{sec:conclusion}

We present RECAL for unsupervised provenance-based intrusion detection through relation-balanced masked graph learning and relation-calibrated anomaly detection. Across three DARPA E3 scenarios, RECAL achieves the highest F1 among the compared methods while maintaining low false positive rates. Future work will explore adaptive relation calibration to address concept drift in long-running deployments.

\clearpage
\section{Compliance with Ethical Standards}
This study uses publicly available DARPA E3 system-audit datasets and involves no human or animal subjects. No ethical approval was required.

\bibliographystyle{IEEEbib}
\urlstyle{same}
\bibliography{references}

\end{document}